\documentclass{elsart5p}

\usepackage{graphics}
\usepackage{graphicx}
\usepackage{amssymb}

\begin{document}

\begin{frontmatter}



\title{Searching for zeroes: unconventional superconductors in a magnetic field}
%

\author{I. Vekhter,}
\author{A. Vorontsov}

\address{Department of Physics and Astronomy, Louisiana State University, Baton Rouge, LA 70803, USA}


\begin{abstract}
We review the results of the microscopic approach to the
calculation of the anisotropy in the specific heat in
unconventional superconductors under rotated field. Treating
vortex scattering on equal footing with the energy shift we find
that the electronic specific heat may have minima or maxima when
the field is aligned with the nodes, depending on the temperature
and field range. We discuss the influence of the paramagnetic
limiting and Fermi surface shape on the location of the inversion
line.
\end{abstract}


\end{frontmatter}

\section{Introduction}
One of the practical steps that help determine the origin of
superconducting pairing in novel materials is the determination of
the shape of the energy gap. There are few techniques that
directly probe the location of the zeroes (nodes) of the
superconducting gap in the bulk of unconventional superconductors.
Prominent among them are measurements of the specific heat and the
thermal conductivity under a rotated magnetic field. Here we
review recent progress on theoretical underpinning of these
experimental techniques, and their implication for realistic
materials.

The basic idea underlying the measurements is based on the
understanding that in superconductors with zeroes of the gap on
the Fermi surface (FS) the unpaired quasiparticles have momenta
close to those of the nodal directions. Specific heat (thermal
conductivity) measure entropy (its transport), and are therefore
insensitive to the condensate of Cooper pairs. Consequently, if an
experiment selectively probes excitations over a region of the FS,
it is able to help locate the gap nodes. Spectroscopic techniques
(e. g. ARPES) can resolve the gap structure in quasi-2D systems
with large gap; however, in materials with low T$_c$ an
alternative method is needed.

Magnetic field provides such an alternative. The field couples to
the phase of the superconducting order parameter, and drives
supercurrents that, in turn, affect near-nodal quasiparticles. In
the mixed state, when the field is only weakly screened, the
interplay between the supercurrents and single particle
excitations occurs throughout the bulk, and is therefore
accessible experimentally.

In the semiclassical approach, valid at low fields, $H\ll H_{c2}$,
and low temperatures, $T\ll T_c$, the position-dependent superflow
around the vortices, ${\bf v}_s(\bf r)$ shifts the energy of the
unpaired electrons relative to the moving condensate, generating a
finite field-dependent density of the quasiparticles. In the
absence of the field the cost of a single electron excitation with
momentum ${\bf k}$ is $E_{\bf k}=\sqrt{\zeta_{\bf
k}^2+|\Delta_{\bf k}|^2}$, where $\zeta_{\bf k}$ is the band
energy, and $\Delta_{\bf k}$ is the superconducting gap. In the
presence of the Doppler shift due to slowly varying superflow this
energy becomes $E^\prime_{\bf k}=E_{\bf k}-{\bf v}_s(\bf
r)\cdot{\bf k}$, and therefore some previously unoccupied states
drop below the chemical potential and become populated. Since the
Doppler shift outside of vortex cores is smaller than the maximal
gap (supercurrent is below the pairbreaking value), the populated
states are located near the gap nodes in the momentum space.

Pioneered by Volovik to predict that the density of states in a
clean superconductors with lines of nodes varies as $\sqrt H$
\cite{GEVolovik:1993}, this approach was extensively used to
analyse nodal superconductors \cite{CKubert:1998}. Early analysis
aimed at high-T$_c$ cuprates, and started with the assumption of a
cylindrical Fermi surface with vertical line nodes. Vekhter et al.
considered the field rotated with respect to nodes in the
conducting planes, and showed that the density of states (DOS)
varies with the angle of field rotation \cite{IVekhter:1999R}.
Since the superfluid velocity is always in the plane normal to the
applied field, the Doppler shift for the nodal quasiparticles,
${\bf v}_s({\bf r})\cdot{\bf k}_n$, depends on the angle between
the field and the nodal direction, and vanishes when the field is
along the node. Consequently, the DOS was predicted to have a
minimum when the field is aligned with nodes, suggesting the
measurements of the electronic specific heat as a test for the
nodal directions. This behavior was indeed found in the vortex
state of several unconventional superconductors \cite{TPark:2003}.
The results were qualitatively consistent with the predictions of
the semiclassical theory, even though the experiments were done at
higher $T/T_c\geq 0.1$ and higher $H/H_{c2}$ than the
corresponding range in the cuprates.

Experiments on the anisotropy of thermal conductivity,
$\kappa(H,T)$, under a rotated field continued for over 10 years,
and unambiguously showed the superposition of the variation due to
difference in transport normal and parallel to the vortices
(``twofold''), and the additional features due to nodal structure
\cite{YMatsuda:2006}. Theoretical analysis of these measurements
remained incomplete until recently. The Doppler shift approach
leaves the lifetime in the absence of impurities infinite, and
therefore does not account for the scattering on the vortices. The
magnetic field does influence lifetime via the self-consistency
between the DOS and the impurity scattering, so that the field
dependence of $\kappa$ is determined by the competition between
the enhanced density of states and the change in lifetime.
Consequently, it is not clear whether rotating the field through
the nodes, for a fixed direction of the heat current, gives minima
(as the DOS) or maxima (due to reduced scattering). Moreover, the
connection between the local $\kappa({\bf r})$ and the measured
value $\kappa({\bf H})$ for a given field direction is not fully
established, and 'ad hoc' averaging procedures often entirely miss
the twofold anisotropy.

The need for a complete theoretical analysis, accounting for the
superflow beyond semiclassical treatment and addressing the
scattering on the vortices, was emphasized by comparison of the
data on the specific heat and the thermal conductivity anisotropy
in heavy fermion CeCoIn$_5$, where the opposite gap structure was
inferred from the specific heat \cite{HAoki:2004} and thermal
transport \cite{KIzawa:115} measurements. Such a description was
recently developed by us \cite{AVorontsov:2006,AVorontsov:2007},
and here we summarize the salient features of our approach, review
the essential physics, and focus on a new aspect of the analysis,
the effect of the Fermi surface shape.

\section{Microscopic approach}
Our approach is based on the quasiclassical formulation of the
microscopic theory. In the spin and particle-hole space the matrix
propagator, which depends on the direction at the Fermi surface
(FS), $\widehat{\bf p}$ , and the center of mass coordinate, ${\bf
R}$, and the mean field order parameter are
    \begin{equation}
      \widehat g({\bf R}, \widehat{\bf p}; \varepsilon) =
    \left( \begin{array}{cc}
    g & i\sigma_2 f \\
    i\sigma_2 \underline{f} & -g
    \end{array} \right) \,,\,
    \widehat\Delta = \left(
    \begin{array}{cc}
    0 & i\sigma_2 \Delta \\
    i\sigma_2 \Delta^* & 0
    \end{array} \right) \,.
    \label{eq:dg}
    \end{equation}
The equilibrium propagators obey
    \begin{eqnarray}
    [(\varepsilon + \frac{e}{c} {\bf v}_F(\widehat{\bf p}) {\bf A}({\bf R}) )\, \widehat{\tau}_3
    - \widehat\Delta({\bf R}, \widehat{\bf p}) - \widehat\sigma_{imp}({\bf R}; \varepsilon),
    \widehat g({\bf R}, {\widehat{\bf p}}; \varepsilon)]
    \nonumber \\
    + i{\bf v}_F({\widehat{\bf p}}) \cdot \nabla_R \;
    \widehat g({\bf R}, {\widehat{\bf p}}; \varepsilon) = 0 \,,\qquad \label{eq:eil}
    \end{eqnarray}
with the normalization condition $\widehat g^2({\bf
R},\hat{\widehat{\bf p}}; \varepsilon) = -\pi^2 \widehat{1}$. In
Eq.(2) ${\bf v}_F(\widehat{\bf p})$ is the Fermi velocity for the
direction $\widehat{\bf p}$, the vector potential ${\bf A}({\bf
R}$ describes the applied field that we take to be uniform, and
the impurity self-energy, $\widehat\sigma_{imp}({\bf R};
\varepsilon)$, is determined within the self-consistent $t$-matrix
approximation. The order parameter is computed self-consistently
with the separable pairing intercation $V({\widehat{\bf
p}},{\widehat{\bf p}}') = V_0\, \cal{Y}({\widehat{\bf p}}) \,
\cal{Y}({\widehat{\bf p}}')$, where $\cal{Y}({\widehat{\bf p}})$
is the normalized basis function for the angular momentum
representation. We consider quasi-two dimensional (open along $z$)
Fermi surfaces and order parameters with vertical lines of nodes.

We model the vortex state by the extended Abrikosov solution
consisting of superposition of both the lowest and higher Landau
level functions, $\Phi_n(x)$, required for unconventional
superconductors, $\Delta({\bf R})= \sum_n\Delta_n \langle{{\bf
R}}|{n}\rangle$ with
\begin{equation}
  \langle{{\bf R}}|{n}\rangle=\sum_{k_y} C_{k_y}^{(n)} {e^{ik_y\sqrt{S_f} y}
    \over \sqrt[4]{S_f \Lambda^2}} \Phi_n\left( {x-\Lambda^2
    \sqrt{S_f} k_y\over \Lambda \sqrt{S_f}} \right)
    \,.
\end{equation}
For the field applied at an angle $\theta_H$ to the $z$-axis, the
anisotropy factor $S_f = \sqrt{ \cos^2 \theta_H + ({v_{0||}/
v_{0\perp})^2} \sin^2 \theta_H}$, where $v_{0\perp}^2 = 2
\langle{\cal{Y}}^2(\widehat{\bf p}) v^2_{\perp i}(p_z)
\rangle_{FS}$, $v_{0\parallel}^2 = 2 \langle
{\cal{Y}}^2(\widehat{\bf p}) v^2_\parallel(p_z) \rangle_{FS}$,
$\langle\bullet\rangle_{FS}$ denotes the Fermi surface average,
$v_\parallel$ is the projection of ${\bf v}_F$ on the $z$ axis,
and $v_{\perp i}$ with $i=x,y$ is its projection on the axes
normal to $z$.

We use the Brantd-Pesch-Tewordt (BPT) approximation in replacing
the normal (diagonal) part of the propagator with its spatial
average, solve for the off-diagonal components in terms of $g$,
and use the normalization condition to determine the
quasiclassical Green's function. We then compute the density of
states directly, and use the linear response theory in Keldysh
formulation to determine the thermal conductivity. The
justification of the approximation, and the technical details of
the approach are given in Ref. \cite{AVorontsov:2007} for a
rotationally symmetric Fermi surface in the shape of a corrugated
cylinder. Here we briefly review the main results of that work,
discuss its relation to the data on CeCoIn$_5$, and focus on the
influence of the Zeeman field and the shape of the Fermi surface
on the conclusions and the phase diagram.

\section{Results.}

In Refs. \cite{AVorontsov:2006,AVorontsov:2007}we considered a
Fermi surface given by $p_f^2 = p_x^2 + p_y^2 - (r^2 \, p_f^2)
\cos (2 s\, p_z/r^2 p_f)$, with $s=r=0.5$, and the $d$-wave gaps,
${\cal Y}=\sqrt 2\cos 2\phi$, where $\phi$ is the azimuthal angle
at the Fermi surface. The unexpected result of
Ref.\cite{AVorontsov:2006} was the inversion of the anisotropic
profile of the specific heat across much of the phase diagram in
the $T$-$H$ plane. In the low temperature, low field region the
results of the microscopic calculation agreed with that of the
semiclassical approach, and the specific heat for the magnetic
field applied along a nodal direction was lower than that for the
field along a gap maximum. In contrast, upon increasing either the
field or the temperature, the pattern changed and the maximum,
rather than a minimum of the specific heat was found when the
field is aligned with the gap node. For a corrugated cylindrical
Fermi surface reminiscent of that of CeCoIn$_5$, the inversion
occurred already at $T/T_c\sim 0.2$ at low fields, and
$H/H_{c2}\sim 0.5$ at low temperature. Therefore over a very large
part of the $T$-$H$ phase diagram, away from the semiclassical
low-energy regime, the theory predicts the anisotropy of the
specific heat opposite to that expected at low fields. As a result
we suggested that the measurements in CeCoIn$_5$, performed at
moderate $H$ and $T$, were consistent with the $d_{x^2-y^2}$
symmetry. Our analysis of the thermal conductivity accounted for
the first time both for the twofold anisotropy due to vortex
scattering and for the structure due to nodes, and supported the
$d_{x^2-y^2}$ symmetry of the order parameter.

\begin{figure}[t]
\begin{center}
\includegraphics[angle=0,width=0.45\textwidth]{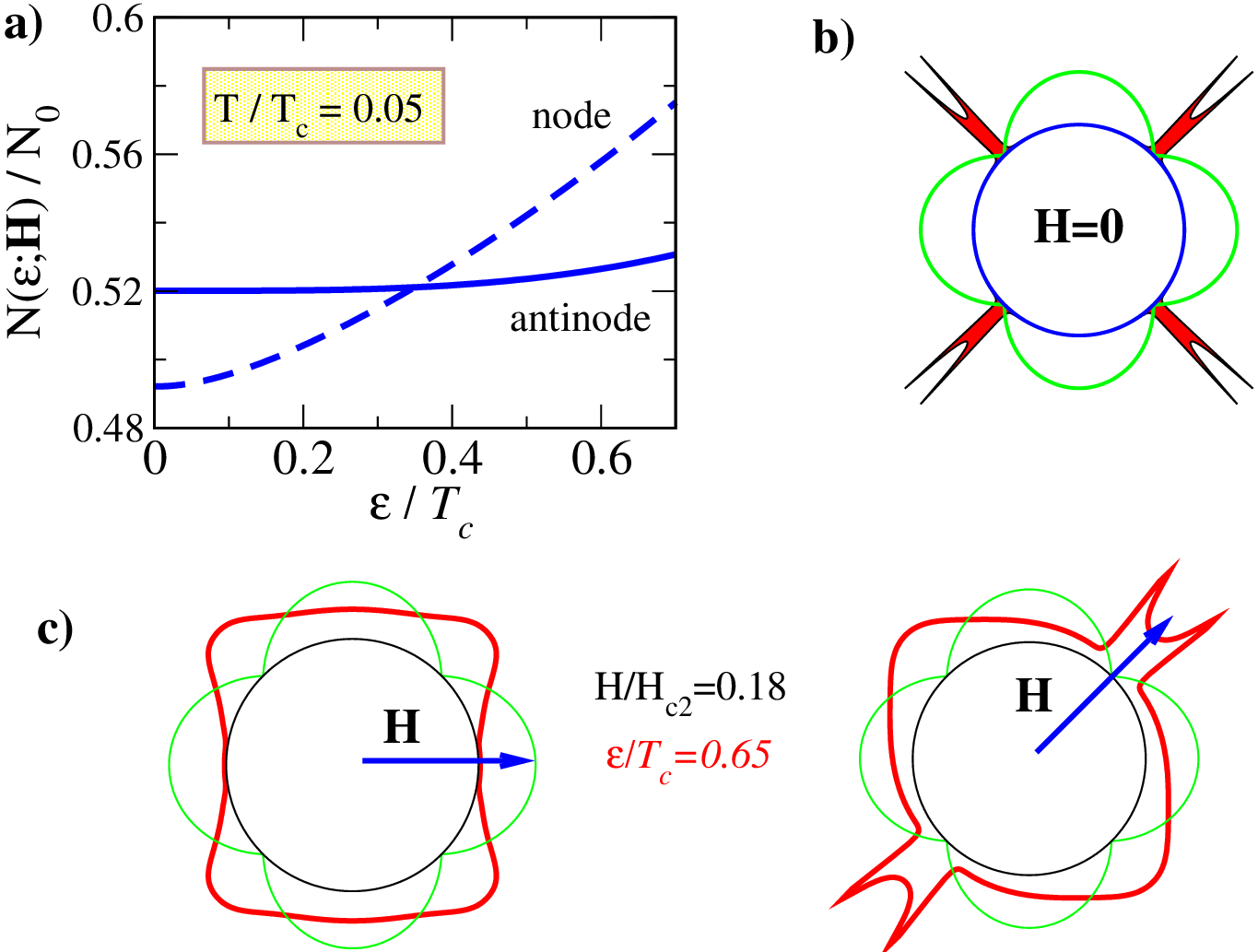}
\end{center}
\caption{Origin of the inversion of the specific heat anisotropy.
a) Low-energy density of states for $H/H_{c2}=0.18$ and field in
the nodal and antinodal directions; b) Typical angle-resolved DOS
(red) in zero field; c) Corresponding angle resolved DOS (red)
under applied field. The order parameter profile is shown for
comparison (green).} \label{fig1}
\end{figure}

The inversion is traced to the effect of vortex scattering as
shown in Fig. 1. At low $T$ the specific heat is related to the
DOS $N(\widehat{\bf p},\varepsilon)/N_0=-{\rm Im} g(\widehat{\bf
p},\varepsilon)$ via
\begin{equation}
  C(T,{\bf H}) = \frac{1}{2}  \int\limits^{+\infty}_{-\infty}
    \; d\varepsilon \; \frac{\varepsilon^2 \; N(T, {\bf H}; \varepsilon)}{T^2
    \cosh^2(\varepsilon/2T)} \,
\end{equation}
and is most sensitive to the DOS at $\varepsilon\sim 2T$.
Inversion of the $C(T,{\bf H})$ anisotropy at moderate $T$ for a
given $H$ implies that the DOS anisotropy for the nodal and
antinodal field directions changes sign, see Fig. 1a): at low
(moderate) energies $N({\bf H}\|node, \varepsilon)<N({\bf
H}\|antinode, \varepsilon)$ ($N({\bf H}\|node, \varepsilon)>N({\bf
H}\|antinode, \varepsilon)$). Consider the angle-resolved density
of states and recall that in the absence of the field the major
contribution to $N(\varepsilon\ll T_c)$ is from the coherence
peaks located at momenta $\widehat{\bf p}_\varepsilon$ such that
$\varepsilon=\Delta_0{\cal Y}(\widehat{\bf p}_\varepsilon)$, close
to the nodal directions, see Fig. 1b). Vortex scattering is weak
for the quasiparticles traveling nearly parallel to the field, and
strong for those with the momenta at moderate to large angles with
respect to ${\bf H}$. Consequently, when the field is applied
along a nodal direction, it largely preserves the spectral weight
in the coherence peaks aligned with the field, while exciting
unpaired quasiparticles at the other two nodes. In contrast, the
field applied along the gap maxima induced more low energy states,
but smears out the coherence peaks at moderate $\varepsilon$,
transferring their spectral weight to other energies, see Fig.
1c). As a result, the DOS anisotropy changes sign at a finite
$\varepsilon^\star$, which was determined numerically.

While the results were very suggestive of the $d_{x^2-y^2}$
symmetry in CeCoIn$_5$, specific details pertinent for this
material left some questions open. First, the upper critical field
in CeCoIn$_5$ is Pauli-limited \cite{ABianchi:2002}, and spin
splitting of the Fermi surface may be important. We considered its
effect by including the Zeeman term, $v_z=g\mu_B \sigma\cdot{\bf
H}$ with $g=2$ into the semiclassical equations, and writing
explicitly the equations for the two spin components, which are
independent. On the other hand, the self-consistency involves spin
summation. We found that the inclusion of Zeeman splitting
modifies the location of the inversion lines in the phase diagram,
but does not affect the main conclusion, namely, the inversion of
the anisotropy at moderately low $H$ and $T$, see Fig. 2. The
effect of the paramagnetic limiting on the phase diagram is
stronger at high fields and low temperatures due to
``isotropization'' of the upper critical field: the orbital
limiting field differs for the nodal and antinodal orientation of
the field, while the Pauli limiting field is
direction-independent. Since in practice the measurements of the
anisotropy are carried out at moderate fields, our conclusions
remain substantially unmodified.
\begin{figure}[t]
\begin{center}
\includegraphics[angle=0,width=0.45\textwidth]{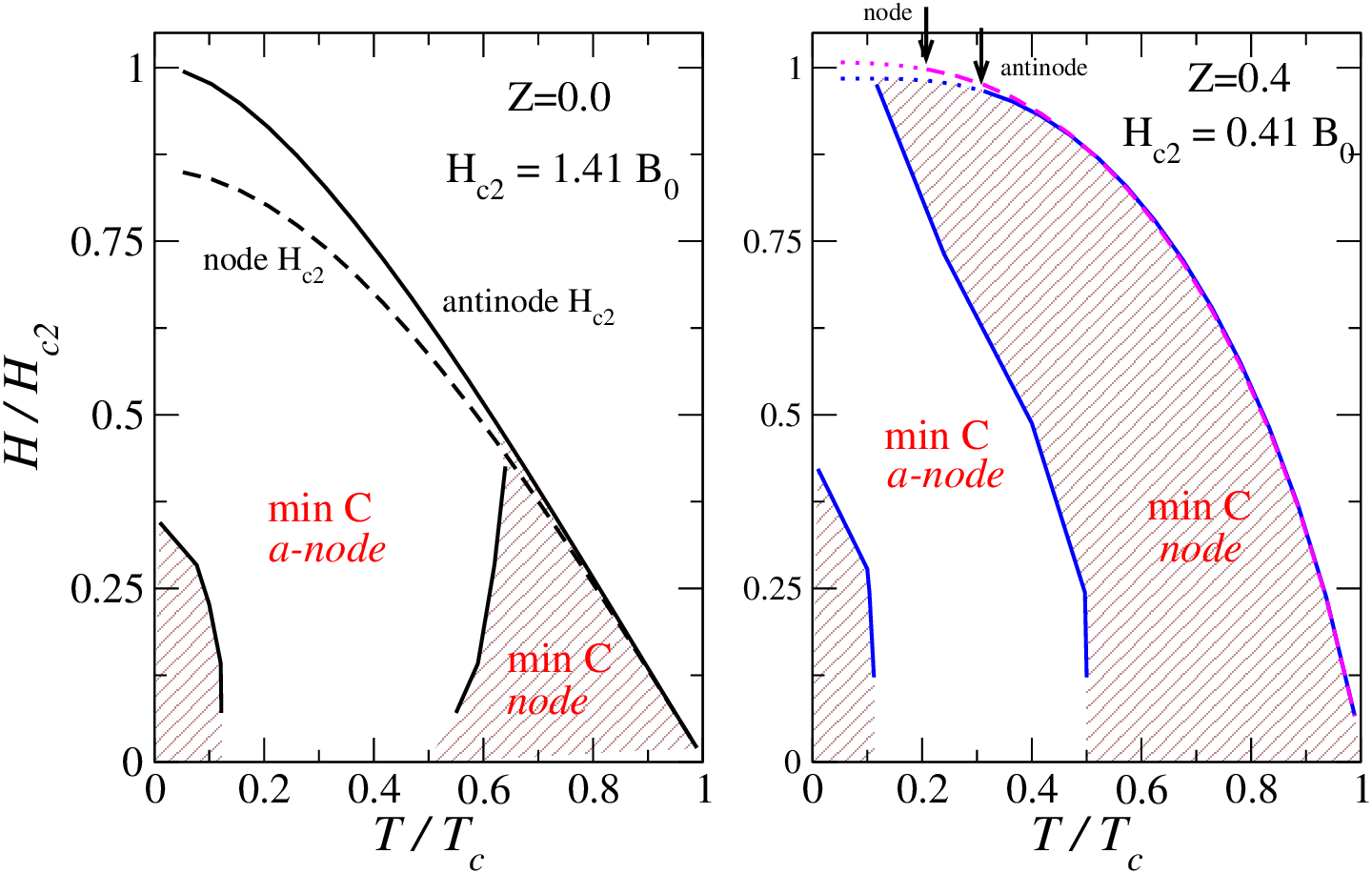}
\end{center}
\caption{Comparison of the phase diagram for the specific heat as
a function of Zeeman splitting. In the shaded (unshaded) area
$C(T,{\bf H})$ has a minimum for the field along the nodal
(antinodal) direction. Here $B_0 = (ch/2e)/2\pi\xi_0^2$ with
$\xi_0 = \hbar v_F /2\pi T_c$ is a measure of the orbital critical
field. Parameter Z = $\mu B_0/2\pi T_c$ characterizes the strength
of the Zeeman coupling. The arrows in the right panel indicate
onset of the first order transition occurring for $Z\geq 0.35$.}
\label{fig2}
\end{figure}

Second, the semiclassical analysis suggests that the local
curvature of the Fermi surface around the nodal points affects the
filling of the near-nodal states due to the field: for a fixed
position and superflow ${\bf v}_s({\bf r})$, the variation of the
Fermi velocity ${\bf v}_F(\widehat{\bf p})$ with direction
$\widehat{\bf p}$ near the node affects the Doppler shift.
Moreover, in borocarbide superconductors some of the features of
the anisotropy profile were suggested to be due to the nested
parts of the Fermi surface \cite{MUdagawa:2005}. Consequently, we
explored the influence of the Fermi surface geometry on our
conclusions.

We consider a model two-dimensional Fermi surface given by
$p_F^2=p_x^2+p_y^2+a^2(p_x^4+p_y^4)$, and chose $ap_F=2$. In this
case the directions along the axes and $p_x=p_y$ are inequivalent,
and therefore the map of the specific heat anisotropy in the
$T$-$H$ plane depends on whether the order parameter is of the
$d_{x^2-y^2}$ or $d_{xy}$ type, with nodes along [110] or [100].
We consider both cases, and model the order parameter by
$\Delta(p_x,p_y)=\Delta_0(p_x^2-p_y^2)$ and
$\Delta(p_x,p_y)=\Delta_0 (2p_xp_y)$ respectively, with $p_x,p_y$
at the FS. Therefore the nodes are located either in the ``flat''
(low curvature) or in the ``corner'' (high curvature) part of the
Fermi surface.

While it is not possible to carry out the fully self-consistent
calculation in field for a FS with no $c$-axis energy dispersion,
we showed in Ref.\cite{AVorontsov:2007} that solving for
$\Delta_0(T,H=0)$ and then modeling the vortex state by the lowest
Landau Level with $\Delta(T,H)=\Delta_0(T,0)\sqrt{1-H/H_{c2}}$
gives results in semi-quantitative agreement with the
self-consistent calculation for a corrugated FS. This is also seen
from comparing the  left panels of Fig.~2 (self-consistent,
corrugated FS) and  Fig.~3 (2D FS).  We show the results in
Fig.~3. It is clear from that the location of the anisotropy
inversion lines is sensitive to the shape of the Fermi surface.

If the nodes of the gap are at the regions of high FS curvature,
the ``semiclassical'' regime of minima of the specific heat for
the field along the nodes is suppressed. Now even for the field
along the node the near-nodal quasiparticles have the velocity at
a substantial angle to the field direction, and are efficiently
scattered by the vortices; hence the anisotropy crossover line is
pushed to ultra-low fields, where we cannot detect it numerically.
Over a large part of the phase diagram the minima of the specific
heat correspond to the field along the gap maxima. Similarly to
the rotationally symmetric Fermi surface, the anisotropy is
reversed at low $T$ and high fields, but in this region near the
transition line the amplitude of the oscillations is small, and
hence is of little experimental relevance.

In contrast, flattening of the near-nodal Fermi surface slightly
extends the ``semiclassical'' region. In this geometry when the
field is applied along a nodal direction, the Doppler shift nearly
vanishes not just at the nodal point where ${\bf k}_n$ is normal
to the superfluid velocity, ${\bf v}_s$, but over the entire
near-nodal region. Moreover, the scattering of the quasiparticles
on vortices in this region is small, since their velocity is
nearly parallel to the field. As a result, semiclassical results
have a somewhat extended regime of validity.

In CeCoIn$_5$ the FS is not rotationally symmetric
\cite{YHaga:2001}, and likely locations of the nodes pass through
regions of high curvature. Therefore the inversion line is likely
to be at lower $T,H$, than a cylindrical FS model predicts.

\begin{figure}[t]
\begin{center}
\includegraphics[angle=0,width=0.45\textwidth]{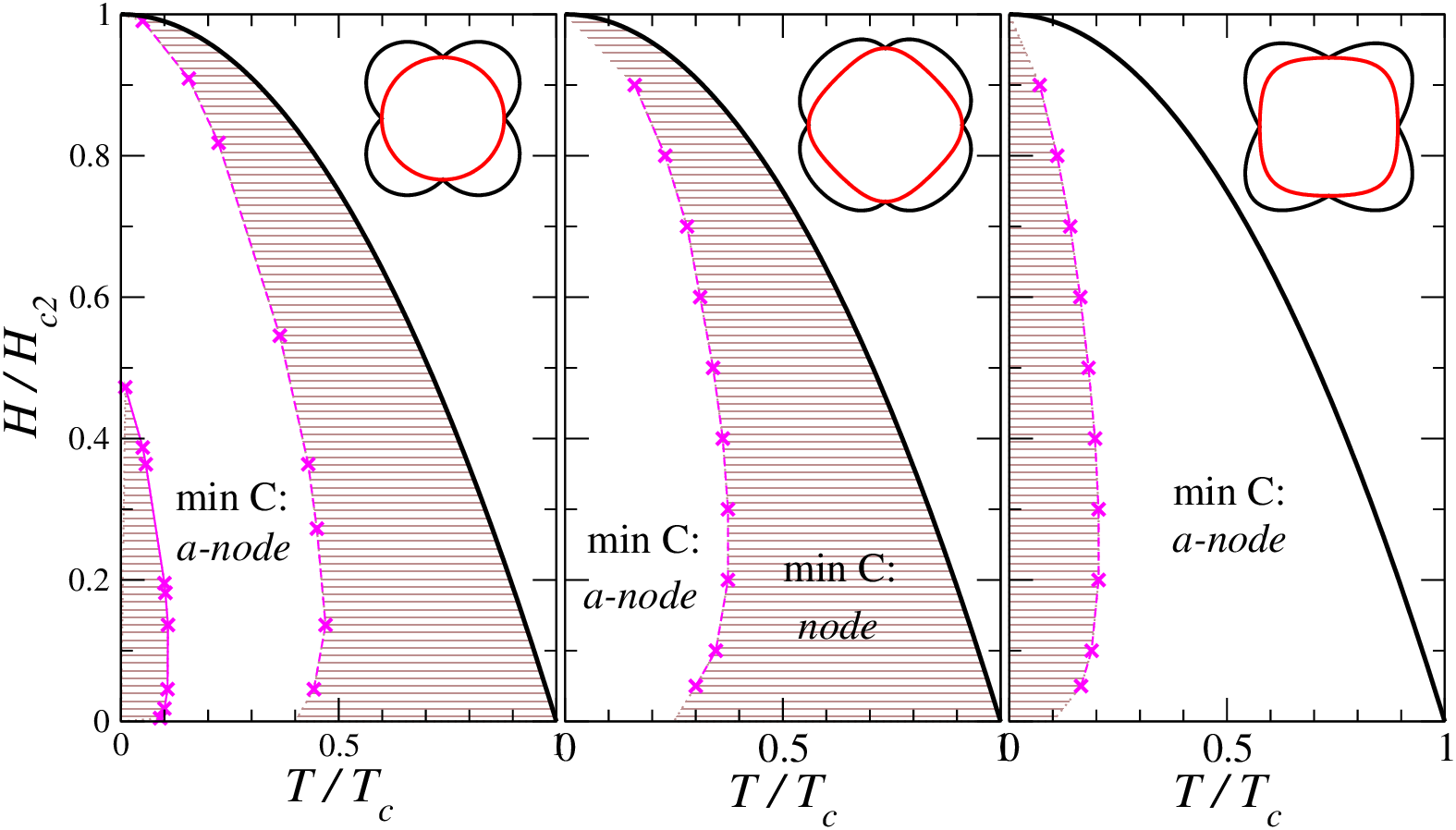}
\end{center}
\caption{Effect of the Fermi surface shape on the anisotropy map
of the specific heat. Left panel: cylindrical Fermi surface;
middle panel: nodes in the region of high curvature; right panel:
nodes in the region of low curvature. In the shaded (unshaded)
area $C(T,{\bf H})$ has a minimum for the field along the nodal
(antinodal) direction. } \label{fig3}
\end{figure}

\section{Summary}

We reviewed here recent progress in microscopic calculations of
the electronic specific heat of unconventional superconductors
under a rotated magnetic field, and its connection with the
experimental efforts to use the method for determination of the
nodal directions. The approach uses the Brandt-Pesch-Tewordt
approximation, which is nearly exact at high fields, and likely
overestimated the effect of vortex scattering at low fields. The
major difference with the results of the semiclassical
approximation is the inversion of the anisotropy of the density of
states and the specific heat due to the competition between the
Doppler energy shift and spectral weight redistribution resulting
from scattering on the vortices. The low-to-moderate $T$ and $H$
location of the inversion line makes it of experimental relevance,
and we showed that the field and temperature range of this
inversion is weakly dependent of paramagnetic effects, but is
sensitive to the shape of the Fermi surface. A more complete
analysis of the Fermi surface effects and Zeeman splitting,
including the behavior of the thermal conductivity, will be
published elsewhere.

We benefited from enlightening discussions with Y. Matsuda and T.
Sakakibara. This research was supported in part by the Louisiana
Board of Regents.

\end{document}